\documentclass[11pt]{article}
\pdfoutput=1 
\usepackage{amsmath,epsf,amssymb,latexsym,enumerate,cite,shadow,array,color,bbm,slashed}
\usepackage{jheppub}

\csname @addtoreset\endcsname{equation}{section}

\DeclareFontFamily{U}{rsf}{}
\DeclareFontShape{U}{rsf}{m}{n}{
  <5> <6> rsfs5 <7> <8> <9> rsfs7 <10-> rsfs10}{}
\DeclareMathAlphabet\Scr{U}{rsf}{m}{n}

\def\pplogo{\vbox{\kern-\headheight\kern -29pt
\halign{##&##\hfil\cr&{
\ppnumber}\cr\rule{0pt}{2.5ex}&\ppdate\cr}
}}
\makeatletter
\def\ps@firstpage{\ps@empty \def\@oddhead{\hss\pplogo}%
  \let\@evenhead\@oddhead 
}

\makeatletter
\renewcommand\section{\@startsection {section}{1}{\z@}%
                                   {-3.5ex \@plus -1ex \@minus -.2ex}%
                                   {2.3ex \@plus.2ex}%
                                   {\normalfont\large\bfseries}}
\renewcommand\subsection{\@startsection{subsection}{2}{\z@}%
                                   {-3.25ex\@plus -1ex \@minus -.2ex}%
                                   {1.5ex \@plus .2ex}%
                                   {\normalfont\normalsize\bfseries}}

\def\rmi{{\rm i}}
\def\rmd{{\rm d}}

\newcommand{\be}{\begin{equation}}
\newcommand{\ee}{\end{equation}}
\newcommand{\ba}{\begin{eqnarray}}
\newcommand{\ea}{\end{eqnarray}}
\newcommand{\lp}{\left(}
\newcommand{\rp}{\right)}
\newcommand{\ls}{\left[}
\newcommand{\rs}{\right]}

\newcommand{\w}{\wedge}

\newcommand{\N}{\mathcal{N}}

\newcommand{\cF}{\mathcal{F}}

\def\ib{{\bar \imath}}

\newcommand{\cG}{{\cal G}}

\newcommand{\cN}{{\cal N}}

\def\be{\begin{equation}}
\def\ee{\end{equation}}

\usepackage{color}

\title{String Theory Origin of Constrained Multiplets}

\author[a]{Renata Kallosh,}
\author[b]{Bert Vercnocke,}
\author[c]{Timm Wrase}

\affiliation[a]{Department of Physics, Stanford University, 382 Via Pueblo, Stanford, CA 94305, USA}
\affiliation[b]{Institute for Theoretical Physics, University of Amsterdam, Science Park 904, Postbus 94485, 1090 GL Amsterdam, The Netherlands}
\affiliation[c]{Institute for Theoretical Physics, TU Wien, Wiedner Hauptstr. 8-10, A-1040 Vienna, Austria}

\emailAdd{kallosh@stanford.edu}
\emailAdd{bert.vercnocke@uva.nl}
\emailAdd{timm.wrase@tuwien.ac.atfirst@one.univ}

\abstract{We study the non-linearly realized spontaneously broken supersymmetry of the (anti-)D3-brane action in type IIB string theory. The worldvolume fields are one vector $A_\mu$, three complex scalars $\phi^i$ and four 4d fermions $\lambda^0$, $\lambda^i$. These transform, in addition to the more familiar $\N=4$ linear supersymmetry, also under 16 spontaneously broken, non-linearly realized supersymmetries. We argue that the worldvolume fields can be packaged into the following constrained 4d non-linear ${\cal N}=1$ multiplets: four chiral multiplets $S$, $Y^i$ that satisfy $S^2=SY^i=0$ and contain the worldvolume fermions $\lambda^0$ and $\lambda^i$; and four chiral multiplets $W_\alpha$, $H^i$ that satisfy $S W_\alpha=S \bar D_{\dot \alpha} \bar H^\ib=0$ and contain the vector $A_\mu$ and the scalars $\phi^i$. We also discuss how placing an anti-D3-brane on top of intersecting O7-planes can lead to an orthogonal multiplet $\Phi$ that satisfies $S(\Phi-\bar \Phi)=0$, which is particularly interesting for inflationary cosmology.}

\begin{document}
\maketitle
\flushbottom
%
%
%
%
%
%
%
%


\section{Introduction}

In attempts to describe the observable universe one finds that non-linearly realized supersymmetry in string theory and supergravity is a helpful tool. Good examples are de Sitter vacua in 4-dimensional $\N=1$ supergravity that describe dark energy. Such de Sitter supergravities, without scalar fields, were recently constructed in \cite{Bergshoeff:2015tra} by promoting the Volkov-Akulov (VA) model with global non-linear supersymmetry \cite{Volkov:1973ix} to supergravity with local supersymmetry. The corresponding chiral nilpotent Goldstino multiplet $S$ \cite{rocek,Komargodski:2009rz}  constrained by the condition 
\be
S^2=0\,,
\ee 
is equivalent to the VA theory via a non-linear local field redefinition \cite{Kuzenko:2010ef}. This kind of multiplet is present on a D3-brane as well as on an anti-D3-brane \cite{Ferrara:2014kva} in a gauge of the local fermionic $\kappa$-symmetry where the Wess-Zumino term vanishes \cite{Aganagic:1996nn,Bergshoeff:1997kr}. This is in agreement with the fact \cite{Kallosh:1997aw} that in the absence of scalars and vectors the D3-brane as well as the anti-D3-brane have a Volkov-Akulov type action. 

In the context of the KKLT construction of de Sitter vacua in string theory \cite{Kachru:2003aw} a different choice of $\kappa$-symmetry fixing, compatible with an anti-D3-brane placed on the top of an orientifold plane, is useful.
In this more appropriate gauge, the analysis of \cite{Kallosh:2014wsa, Kallosh:2015nia} allows an interpretation of the KKLT construction of de Sitter vacua within a four-dimensional supersymmetric theory. This builds upon early investigations on supersymmetry breaking in string theory \cite{Pradisi:1988xd}, and it transpires that the low energy effective action for an anti-D3-brane on top of an O3-plane in a supersymmetric GKP background \cite{Giddings:2001yu} is just the VA action \cite{Kallosh:2014wsa}. Such an anti-D3-brane on top of an O3-plane can arise in many warped throats  \cite{Kallosh:2015nia} (including the Klebanov-Strassler throat \cite{Klebanov:2000hb}).

An early argument that D-branes have to be associated with spontaneously broken, rather than explicitly broken supersymmetry, was given by J.\ Polchinski in his book \cite{Polchinski:1998rr} (brane actions as effective descriptions of partially spontaneously broken supersymmetry even go back to \cite{Hughes:1986fa}). A more specific prediction with regards to anti-D3-branes was presented in a series of papers by S.\ Kachru and his collaborators in \cite{Shamit}. It was argued there that the system must be viewed as spontaneous breaking of supersymmetry, because it can tunnel to a supersymmetric vacuum. Recent holographic studies point towards spontaneous breaking as well \cite{Bertolini:2015hua}. The nilpotent ${\cal N}=1$ multiplet is the beginning of the explicit realization of the expectation in \cite{Polchinski:1998rr}, \cite{Shamit} and it is natural to look at anti-D3-branes to find other constrained superfields which transform under the non-linearly realized spontaneously broken $\N=1$ supersymmetry. This is the goal of this paper.

In \cite{Vercnocke:2016fbt} it was already shown, using the same type of gauge-fixing as in \cite{Kallosh:2014wsa}, that, in absence of vectors and scalars on the brane,  in addition to a 4d nilpotent multiplet $S^2=0$, there is also a triplet of `scalar-less' chiral multiplets $Y^i$ present, that satisfies $SY^i=0$, $i=1,2,3$. It was also conjectured there that the worldvolume vector field $A_\mu$ and the transverse complex scalars $\phi^i$ can be packaged into constrained multiplets that satisfy $SW_\alpha=0$ and $S D_{\dot \alpha} \bar H^\ib=0$, respectively.\footnote{The coupling of $S$ and $Y^i$ to supergravity was studied in \cite{Dall'Agata:2015zla}. A general approach to couple constrained superfields to gravity was developed recently in \cite{Ferrara:2016een} and a universal way of obtaining constrained multiplets was derived in \cite{Dall'Agata:2016yof}.}

The purpose of this paper is to establish that  indeed  all world-volume fields on the anti-D3-brane, which under linearly realized supersymmetry represent a 4d ${\cal N}=4$ vector multiplet, a vector $A_\mu$, 6 real scalars $\phi^I_r$ and 4 spinors $\lambda^A$, can be packaged into constrained ${\cal N}=1$ superfields with a non-linearly realized supersymmetry.  For this purpose we will use the $\kappa$-symmetry gauge fixing where the Wess-Zumino term vanishes \cite{Aganagic:1996nn,Bergshoeff:1997kr}.\footnote{A related class of $\kappa$-symmetry gauges for a D3-brane in a supergravity backgrounds was studied in \cite{Grana:2002tu}.} This leads to the gauge-fixed Dirac-Born-Infeld-Volkov-Akulov action which is the same for the D3-brane and for the anti-D3-brane. It has 16 linear supersymmetries, the usual linear ${\cal N}=4$ supersymmetry preserved by an (anti-) D3-brane in flat space, and another 16 non-linear supersymmetries of Volkov-Akulov type that are spontaneously broken. 

We are, in particular,  motivated here by the issue of cosmology, where constrained superfields proved to be very useful  \cite{Antoniadis:2014oya, Ferrara:2014kva,Dall'Agata:2014oka}, see also the reviews \cite{Ferrara:2016ajl}. We 
 would like to find out which particular  ${\cal N}=1$ superfields, in addition to $S^2=0$, live on a D-brane. For example, it would be nice to see if the orthogonal nilpotent inflaton superfield \cite{Komargodski:2009rz,Ferrara:2015tyn}, of the type $S B=0$ with $B=\frac1{ 2\rmi}( \Phi-\bar \Phi)$ and $B^3=0$ lives on a D-brane and might allow string theory realizations of the models of inflation without sGoldstino, sinflaton and inflatino \cite{Carrasco:2015iij}. We argue here that such an orthogonal nilpotent superfield can indeed arise on an anti- D3-brane but we leave a detailed study for the future.

The outline of the paper is as follows: In section \ref{sec:N=1truncations} we discuss a variety of different truncations of the D3-brane action that have already appeared in the literature. These provide strong support for our identification of the worldvolume degrees of freedom with the above constrained chiral multiplets. In section \ref{sec:3brane} we review the D3-brane action and show that the non-linear supersymmetry transformations of the worldvolume fields take, after appropriate field redefinitions, a standard non-linear form. We conclude and summarize our findings in section \ref{sec:conclusion}. Appendix \ref{sec:beta} contains some technical details of the calculations performed in section \ref{sec:3brane} and appendix \ref{app:nonlinear} contains more information on non-linear realizations of supersymmetry and on constrained superfields.

\section{Evidence from different \texorpdfstring{$\N=1$}{} truncations}\label{sec:N=1truncations}

In this section we gather evidence from the literature to support our claim that the worldvolume fields on an anti-D3-brane that spontaneously breaks $\N=1$ supersymmetry give rise to constrained $\cN=1$ chiral multiplets $S,Y^i,H^i,W_\alpha$ that satisfy
\be
S^2=0\,,\qquad S Y^i=0\,,\qquad S \bar{D}_{\dot{\alpha}} \bar H^\ib =0\,, \qquad S W_\alpha=0\,.\label{eq:constraints}
\ee
We will do so in three different $\cN=1$ truncations, for which we also provide an interpretation in terms of ten-dimensional string theory models. These truncations are simplified models of a D3-brane in flat space in ten dimensions, which breaks supersymmetry spontaneously to 16 linear and 16 non-linear supersymmetries. The worldvolume fields that transform under these supersymmetries are a vector field $A_\mu$, three complex scalars $\phi^i$, $i=1,2,3$, and four spinors $\lambda^0$, $\lambda^i$. The three truncations discussed in this section have a nilpotent Goldstino $S$ obeying $S^2=0$, supplemented by either
\begin{enumerate} 
\item a non-linearly transforming vector field constrained by $SW_\alpha=0$,
\item fermions described by $S Y^i=0$, or
\item a transverse scalar from the constraint $S \bar{D}_{\dot \alpha} \bar H=0$.
\end{enumerate}

\subsection{Goldstino plus vector} \label{ss:only_SU(3)singlets}
The ${\cal N}=1$ supersymmetric Born-Infeld action of a 3-brane in four dimensions  is the $\cN=1$ supersymmetrization of the bosonic Born-Infeld action \cite{Cecotti:1986gb}
\begin{equation}
{\cal S} = - \int \rmd^4 x \sqrt{-\det (\eta_{\mu\nu} + F_{\mu\nu})}\,,\qquad \mu, \nu=0,1,2,3\,.
\end{equation}
It has a hidden supersymmetry that is non-linearly realized. One can understand the action as the spontaneous breaking of $\N=2$ to $\N=1$.

The field content, a vector and a fermion, can be described by chiral ${\cal N}=1$ multiplets $S$ and $W_\alpha$, satisfying the condition \cite{Bagger:1996wp}
\be
W^2 + S(1- \frac 14 \bar D^2 \bar S)=0 \, , \qquad S^2=0\,.
\label{BG}
\ee
This gives rise to a {\it complete supersymmetric Born-Infeld action} \cite{Cecotti:1986gb}. It was shown in \cite{Kuzenko:2005wh} that ${\cal N}=1$ supersymmetric nonlinear electrodynamics has a fermionic action which up to non-linear field transformations is in agreement with VA model.

The $\cN=1$ Lagrangian for the field strength superfield $W_\alpha$ is 
\be
{\cal S} = \int \rmd^2 \theta \, S(W,\bar W)+\int \rmd^2 \theta \, \bar S(W,\bar W)\,.
\ee 
One can understand $W_\alpha$ and $S$ to be the components of an $\N=2$ vector multiplet \cite{Rocek:1997hi}; since the $\N=2$ supersymmetry is spontaneously broken one finds that the $\cN=2$ vector multiplet is constrained by \eqref{BG}. 

Since $W_\alpha W^2=0$ and $W^2 W^2=0$, the supersymmetric  Born-Infeld action  can be described using one nilpotent field and an orthogonal vector field strength chiral superfield $W_\alpha$
\be
SW_\alpha=0 \, , \qquad S^2=0\,.
\ee
See also \cite{Ferrara:2014oka,Ferrara:2015exa} where this fact was discussed recently.  Note that the condition $SW_\alpha=0$ is a consequence of the one in \eqref{BG} but not vice versa.

Many earlier studies of nonlinear supersymmetry and duality can be found in \cite{Kuzenko:2000uh, Carrasco:2013qia}, including hidden supersymmetries. It has been argued in \cite{Carrasco:2013qia}  that this hidden supersymmetry happens if and only if there is a Born-Infeld dependence on the Maxwell field strength and a Volkov-Akulov dependence on the Goldstino, up to local nonlinear field redefinitions. The relation to duality  for the ${\cal N}=2$ superfield action with manifest ${\cal N}=2$ supersymmetry and hidden ${\cal N}=2$ supersymmetry at the level up to order ${\cal W}^{10}$, where ${\cal W}$ is an  ${\cal N}=2$ vector multiplet, was tested numerically and confirmed in \cite{Carrasco:2013qia}. 

There is a  very familiar string theory construction in which this particular 4d supersymmetric theory arises, namely KKLT \cite{Kachru:2003aw}. An anti-D3-brane in the KKLT setup can uplift a supersymmetric AdS vacuum to a dS vacuum. The anti-D3-brane breaks the $\cN=1$ supersymmetry spontaneously. Generically, the only massless fields on the anti-D3-brane are a vector field $A_\mu$, whose gauge invariance forbids a mass term, and the Goldstino that is similarly protected from developing a mass. The scalar fields on the anti-D3-brane as well as the fermion triplet obtain a non-zero mass \cite{McGuirk:2012sb, Bergshoeff:2015jxa} and decouple from the low energy effective action. To make the connection with the above discussed supersymmetric Born-Infeld action fully transparent, we can think of the setup as an anti-D3-brane in a CY$_3$ compactification with fluxes. This CY$_3$ compactification preserves $\cN=2$ supersymmetry in four dimensions. In the presence of the anti-D3-brane four supersymmetries are linearly realized and four are non-linearly realized and spontaneously broken as above. In the full KKLT setup we need to include an orientifold projection, which actually removes the four linearly realized supersymmetries in the presence of the anti-D3-brane, so that we are only left with the four supersymmetries that are non-linearly realized and spontaneously broken.

\subsection{Goldstino plus fermions}\label{ss:only_fermions}

In \cite{Vercnocke:2016fbt}, two of us discussed the truncation of the $\cN =4$ Born-Infeld action where we only keep the 4 fermions $\lambda^0,\lambda^i$ of the 3-brane worldvolume theory:
\be\label{eq:SD3O3flat2}
{\cal S}_{\rm fermions} = -2 \int \tilde{E}^0 \w \tilde{E}^1 \w \tilde{E}^2 \w \tilde{E}^3 \qquad \text{with} \qquad \tilde{E}^\mu = dx^\mu + \sum_{A=0}^3 \bar \lambda^A \gamma^\mu d \lambda^A\,.
\ee
We found that the fermions $\lambda^0,\lambda^i$ can be described by one nilpotent superfield and three orthogonal superfields:
\begin{equation}
 S^2 =0\,,\qquad SY^i = 0\,.\label{eq:constraint_fermion_truncation}
\end{equation}
We want to stress that the spinor components of $S$ and $Y^i$ are not to be exactly identified with $\lambda^0$ and $\lambda^i$. There is a non-linear field redefinition involved $S(\lambda^A,\bar \lambda^A), Y^i(\lambda^A,\bar \lambda^A)$. We refer to \cite{Vercnocke:2016fbt} for more information and only summarize the main points here.

The lowest order action in the fermion expansion is reproduced by the following K\"ahler and superpotential for $S$ and $Y^i$
\begin{equation}
K =c  S \bar S + \delta_{i\ib} Y^i \bar Y^{\ib}\,, \qquad W = f S +  h_{ij}  Y^iY^j\,,\label{eq:Lleading}
\end{equation}
where $c \in \mathbb{R}$ and $f, h_{ij} \in \mathbb{C}$. The coefficient $c$ is needed to give a  canonical kinetic term for the spinors after the field redefinition $S(\lambda^A,\bar \lambda^A)$. The parameter $f$ controls the size of the supersymmetry breaking and $h_{ij}$ is the mass matrix $m_{ij} = h_{ij}$ for the fermions $\lambda^i$.

Imposing the constraints in equation \eqref{eq:constraint_fermion_truncation} leads to higher order terms in the fermions that also appear in the 3-brane action. In \cite{Vercnocke:2016fbt}, it was shown that in order to get the entire 3-brane action one would have to demand invariance under an enhanced non-linear $\cN=4$ supersymmetry and not just the residual $\cN=1$ supersymmetry, since the later is not sufficient to obtain terms that are quartic or higher in the spinors $\lambda^i$, $i=1,2,3$.

The string theory setup that gives rise to this truncation with only spinors is an anti-D3-brane on top of an O3-plane (see for example \cite{Sugimoto:1999tx, Antoniadis:1999xk, Uranga:1999ib}). The corresponding anti-D3-brane action in equation \eqref{eq:SD3O3flat2} was derived in \cite{Kallosh:2014wsa}. It was also shown in \cite{McGuirk:2012sb, Bergshoeff:2015jxa} that an anti-D3-brane on top of an O3-plane in a non-trivial flux background of the GKP type \cite{Giddings:2001yu} can have mass terms for the spinors. In particular, if the background fluxes are imaginary self-dual and of type (2,1) so that they by themselves would preserve a linearly realized $\cN=1$ supersymmetry, then the fermion triplet $\lambda^i$ obtains a mass proportional to the flux quanta. The anti-D3-brane action in this background is only known to quadratic order in the fermions and agrees (to this order) with the action in equation \eqref{eq:SD3O3flat2}, if we identify $h_{ij}$ with the fermionic mass matrix $m_{ij}$ (see \cite{Bergshoeff:2015jxa} for details).

\subsection{Goldstino plus complex scalar}\label{ss:only_SU(3)triplets}

One can realize another $\cN=1$ truncation of the three-brane action, such that the only non-zero fields are one complex scalar and one fermion. These fields describe the $\N=1$ supersymmetric Nambu-Goto action of a 3-brane in a 6-dimensional spacetime \cite{Hughes:1986fa}, whose bosonic action is
\begin{equation}
{\cal S} = - \int \rmd ^4 x \sqrt{-\det (\eta_{\mu\nu} + \partial_\mu \phi_r^L \partial_\nu \phi_{r,L})}\,,
\end{equation}
with $L = 1,2$.

The supersymmetric Nambu-Goto action for the transverse real degrees of freedom $\phi_r^1, \phi_r^2$ and the fermion can again be realized as a spontaneous breaking of $\cN=2$ to $\cN=1$ \cite{Bagger:1997pi,Rocek:1997hi,GonzalezRey:1998kh}. The Goldstone multiplet of this breaking is now not an $\cN=2$ vector multiplet, but an $\cN=2$ tensor multiplet. Its $\cN=1$ components are a chiral nilpotent superfield $S$ and a real linear superfield $G$ obeying the constraint
\begin{equation}
 S = S \bar D^2 \bar S + \frac 12 \bar D^{\dot \alpha} G \bar D_{\dot \alpha } G\,,\label{eq:constraint_linear_superfield}
\end{equation}
from which one derives
\begin{equation}
S {\bar D}_{\dot \alpha} G =0\,,
\end{equation}
since  $S^2=0$.
The Lagrangian can be given in the form ${\cal L}_G = -\tfrac 12  G^2 + {\cal L}^{int}_G $
\be
{\cal L}_G = -\frac 12  G^2  + \bar D^{\dot \alpha} G \bar D_{\dot \alpha } G \,  D^{ \alpha} G  D_{ \alpha } G \, F({\cal G}^2, \bar {\cal G} ^2)\,,
\ee
where $F$ is some function of ${\cal G}^2= (D^{ \alpha} \bar D^{\dot \alpha} G) (D_{ \alpha} \bar D_{\dot \alpha} G)$  and its conjugate $\bar {\cal G} ^2$ and can be obtained by solving the constraint \eqref{eq:constraint_linear_superfield}. To show that this is indeed the supersymmetric Nambu-Goto action of a three-brane in flat space, one adds a Lagrange multiplier $  G (H + \bar H)$ and changes variables to a chiral superfield $H$. The scalar component of $H$ describes the transverse directions to the 3-brane in 6-dimensions. One finds that 
\be
H +\bar H = G + (\bar D_{\dot \beta} G)  X^{\dot \beta}\,,
\ee 
where $X^{\dot \beta}$ depends on $G,\bar G, {\cal G}^2, \bar {\cal G}^2$ and their derivatives. Therefore
 \be
\bar D_{\dot \alpha}( H +\bar H) = \bar D_{\dot \alpha}(G + (\bar D_{\dot \beta} G)  X^{\dot \beta})\,,
\ee 
and since $H$ is chiral, $\bar D_{\dot \beta} H=0$, and $G$ is linear, $\bar D_{\dot \alpha }\bar D^{\dot \beta} G=0$, we find that
 \be
\bar D_{\dot \alpha} \bar H = \bar D_{\dot \alpha}G + (\bar D_{\dot\beta } G)   \bar D_{\dot \alpha} X^{\dot \beta}\,.
\label{useful}\ee
We multiply it by $S$ and get the desired result of a relaxed and a nilpotent multiplet
\begin{equation}
 S {\bar D}_{\dot \alpha} \bar H =0   \,,\qquad S^2=0\,.
\label{rel}
\end{equation}
One can rewrite this equally well as a nilpotent chiral multiplet $S$ and two orthogonal chiral multiplets $H = \Phi_1 + \rmi \Phi_2$ that satisfy the constraints (see \cite{Komargodski:2009rz})
\be
S(\Phi_1- \bar \Phi_1)=S(\Phi_2- \bar \Phi_2)=0\,,\qquad S^2=0\,.
\ee

We present a simple string theory construction that gives rise to such a nilpotent and two orthogonal multiplets.\footnote{We thank Angel Uranga and I\~naki Garc\'ia-Etxebarria for useful discussions of this setup.} We use $z^i$ with $i=1,2,3,$ to denote the three complex transverse direction for an anti-D3-brane in flat ten-dimensional spacetime. We again write the world volume fermions in terms of 4D spinors $\lambda^0$ and $\lambda^i$. Now we follow \cite{Kallosh:2015nia} and do an O7-orientifold projection $\Omega_p {\cal R}_1 (-1)^{F_L}$. We take the geometric action such that it leaves $z^1$ and $z^2$ invariant and maps ${\cal R}_1: z^3\rightarrow -z^3$. If the anti-D3-brane is on top of the O7-plane, at $z^3=0$, then the orientifold projection removes the vector field $A_\mu$, the scalar $\phi^3$, and the two fermions $\lambda^1$ and $\lambda^2$.

Now let us look at another O7-plane with geometric action that only acts non-trivially on $z^2$ as ${\cal R}_2: z^2\rightarrow -z^2$. For an anti-D3-brane at $z_2=0$ it removes the vector field $A_\mu$, the scalar $\phi^2$, and the two fermions $\lambda^1$ and $\lambda^3$. If we do both of these orientifold projections, and place the anti-D3-brane at $z^2=z^3=0$, then we remove everything but the scalar $\phi^1$ and the spinor $\lambda^0$. This setup preserves in 4d $\cN=1+1$ supersymmetry: a linear $\cN=1$ supersymmetry under which the two fields form a chiral multiplet and a non-linearly realized and spontaneously broken $\cN=1$ supersymmetry for which $\lambda^0$ is the Goldstino. As in the case of the 3-brane in a 6-dimensional space, we have a complex scalar $\phi^1=\phi^1_r + \rmi \phi^2_r$ that controls the position in the two real transverse directions.

Embedding this setup into a more complicated space transverse to the anti-D3-brane and/or turning on (2,1) ISD fluxes as in GKP, will break the linearly realized $\N=1$ supersymmetry while the non-linearly realized supersymmetry remains. We expect this to generate a potential for the scalar $\phi^1$, however, the Goldstino $\lambda^0$ is protected. Similarly to the anti-D3-brane in KKLT, where generic fluxes give a mass to all transverse scalars and special fluxes preserve flat directions, we suspect that it is likewise possible to give only a mass to for example the real or imaginary part of $\phi^1$. The low energy effective action would then contain the Goldstino $\lambda^0$ and a real scalar $\phi_r$. These low energy degrees of freedom would then be described by a nilpotent chiral multiplet $S$ and an orthogonal multiplet $\Phi$ that satisfies
\be
S^2=0\,, \qquad S(\Phi-\bar \Phi)=0\,.
\ee
It would be very interesting to make this precise and work out a concrete string compactification that gives rise to such an orthogonal multiplet.

\section{The D3-brane action and its non-linear supersymmetries}
\label{sec:3brane}

We now discuss the D3-brane action in flat space and its $\kappa$-symmetry.\footnote{The case of the anti-D3-brane in flat space is the same up to some irrelevant sign flips.} The brane action is invariant under 16 linearly realized supersymmetries, and 16 non-linear ones \cite{Duff:1991pea}. From the four-dimensional world volume viewpoint, this represents a spontaneous breaking of the $\cN=8$ supersymmetry of flat space to $\cN = 4$. After a review of the D3-brane action, we identify the relevant linear and non-linear supersymmetry transformations of the worldvolume fields. We show that, after field redefinitions, the non-linear transformations take the standard form in equations \eqref{eq:nltrafos} and \eqref{eq:nltrafos2} below, in line with the constraints \eqref{eq:constraints}.  We also explain how the particular truncations of the previous section can be obtained. For a review of the connection between these non-linear transformations and the constraints, we refer to appendix \ref{app:nonlinear}. 

\subsection{The generic D3-brane action and its symmetries}

The $\kappa$-symmetric D3-brane action \cite{Cederwall:1996pv, Aganagic:1996nn, Bergshoeff:1997kr} in a flat
background geometry consists of the Dirac-Born-Infeld-Nambu-Goto term ${\cal S}_{\rm DBI}$ and the Wess-Zumino term ${\cal S}_{\rm WZ}$ with world-volume coordinates $\sigma^{\mu}$, $\mu =0, 1, 2, 3$. It was studied  mostly a couple of decades ago and reviewed recently in \cite{Simon:2011rw}. 

Here we use the more recent analysis and notation of \cite{Bergshoeff:2013pia} and \cite{ Kallosh:2014wsa} (we will however set $\alpha'=1$). The action for a D3-brane in flat space, including all the fermionic terms, is given by
\be\label{eq:action}
{\cal S}^{D3} = {\cal S}_{\text{DBI}} + {\cal S}_{\text{WZ}} = - \int d^4 \sigma \sqrt{-\text{det}(G_{\mu\nu} + \cF _{\mu\nu})} +  \int \Omega_4\,.
\ee
We denote the longitudinal and transverse coordinates as 
\be
X^M = \{ X^m, \phi_r^I \}\,,\qquad M=0,1,\ldots, 9\,, \qquad m=0,1,2,3,\qquad I=1,2,3,4,5,6\,,
\ee
where $m$ refers to the worldvolume coordinates and $I$ to the six real transverse coordinates, which we will often write as three complex directions $\phi^i=\phi_r^{2i-1}+\rmi \phi_r^{2i}$, $i=1,2,3$. The $\phi^i$ are the scalar fields that control the position of the anti-D3-brane. The metric including fermionic terms is given by
\be
G_{\mu\nu} = \eta_{mn} \Pi^m_\mu \Pi^n_\nu + \delta_{IJ} \Pi^I_\mu \Pi^J_\nu\,, \qquad \Pi^m_\mu = \partial_\mu X^m-\bar \theta \Gamma^m \partial_\mu \theta\,, \qquad \Pi^I_\mu = \partial_\mu \phi_r^I - \bar \theta \Gamma^I \partial_\mu \theta\,,
\ee
where $\eta_{mn}$ is the Minkowski metric, $\Gamma^M$ are 10D gamma matrices and $\theta=(\theta^1,\theta^2)$ denotes a doublet of 16 components MW spinors of the same chirality so that $\bar\theta = (\theta_1^T C, \theta_2^T C)$ with $C$ being the 10D charge conjugation matrix. When it is clear from the context, we omit the doublet index. We always omit the spinorial indices.

The Born-Infeld field strength $\cF_{\mu\nu}$ is given by
\be
\cF_{\mu\nu} = F_{\mu\nu} -b_{\mu\nu}\,, \qquad b_{\mu\nu}= \bar \theta \sigma_3 \Gamma_M \partial_\mu \theta \lp \partial_\nu X^M-\frac12 \bar \theta \Gamma^M \partial_\nu \theta \rp - (\mu \leftrightarrow \nu)\,,
\ee
where $F_{\mu\nu} = \partial_\mu A_\nu -\partial_\nu A_\mu$ is the field strength of the worldvolume gauge field $A_\mu$. Lastly, the 4-form $\Omega_4$ is defined via a closed 5-form
\be \label{eq:I5}
I_5 = d \Omega_4 = d\bar \theta \lp \sigma_1 \cF \tilde{\Gamma} + \rmi \sigma_2 \frac{\tilde{\Gamma}^3}{3!} \rp d \theta\,,\qquad \tilde{\Gamma} = \Gamma_M \Pi^M = \Gamma_M(dX^M+\bar\theta \Gamma^M d\theta)\,,
\ee
where wedge products are implicit and the plus sign in the last equation above is explained on page 5 of \cite{Aganagic:1996nn}.

The D3-brane action given above in equation \eqref{eq:action} is invariant under several symmetries (as discussed for example in appendix A of \cite{Bergshoeff:2013pia}). The worldvolume fields transform under global supersymmetry with the parameters $\tilde{\epsilon}^1$, $\tilde{\epsilon}^2$, local $\kappa$-symmetry with the parameters $\kappa^1$ and $\kappa^2$ and under world volume diffeomorphisms parameterized by $\xi^\mu$. The transformation laws are as follows
\ba\label{eq:branetrafos}
\delta \theta &=& \tilde{\epsilon}+ (1 + \Gamma)\kappa+\xi^\mu \partial_\mu \theta\,,\cr
\delta X^M &=& -\bar{\theta} \Gamma^M \tilde{\epsilon} + \bar{\theta} \Gamma^M (1+ \Gamma) \kappa + \xi^\mu \partial_\mu X^M\,,\cr
\delta A_\mu &=& -\bar{\theta} \Gamma_M \sigma_3 \tilde{\epsilon}\  \partial_\mu X^M + \frac16 \bar{\theta} \sigma_3 \Gamma_M \tilde{\epsilon} \ \bar{\theta} \Gamma^M \partial_\mu \theta + \frac16 \bar{\theta} \Gamma_M \tilde{\epsilon}\ \bar{\theta} \sigma_3 \Gamma^M \partial_\mu \theta + \bar{\theta} \sigma_3 \Gamma_M (1+\Gamma)\kappa\ \partial_\mu X^M \cr
&&  -\frac12 \bar{\theta} \sigma_3 \Gamma_M (1+\Gamma) \kappa \ \bar{\theta} \Gamma^M \partial_\mu \theta -\frac12 \bar{\theta} \Gamma_M (1 + \Gamma)\kappa \ \bar{\theta}\sigma_3 \Gamma^M \partial_\mu \theta + \xi^\nu F_{\nu\mu} \,,
\ea
where we omitted the $U(1)$ gauge transformation for $A_\mu$ and the definition of $\Gamma$ is not relevant for us here (but given in eqns. (A.9)-(A.14) of \cite{Bergshoeff:2013pia}).

\subsection{The DBI-VA model}\label{sec:fixsimple}
The $\kappa$-symmetry allows one to remove half of the 32 fermionic degrees of freedom in $\theta$ so that after $\kappa$-fixing we are left with a single physical 10d MW fermion $\lambda$ with 16 components. A particularly simple way of fixing the $\kappa$-symmetry is given by \cite{Aganagic:1996nn}
\be\label{eq:theta0}
(\mathbbm{1}+\sigma_3) \theta = 0 \qquad \Leftrightarrow \qquad \theta^1 = 0\,.
\ee
The physical worldvolume fermion is then $\lambda \equiv \theta^2$. In this gauge the WZ-term is constant and the action substantially simplifies to the Dirac-Born-Infeld-Volkov-Akulov action
\be
{\cal S}^{DBI-VA}=  - \int d^4 \sigma\ls \sqrt{-\text{det}(G_{\mu\nu} + \cF _{\mu\nu})}\rs_{\theta^1=0}\,.
\ee
We can also fix the diffeomorphism invariance by imposing $X^m(\sigma)= \delta^m_\mu \sigma^\mu$. The resulting transformations of the worldvolume fields in this gauge have been worked out in appendix A of \cite{Bergshoeff:2013pia}. Defining the new 16 component MW spinors $\epsilon$ and $\zeta$ via 
\be\label{eq:spinorredef}
\tilde{\epsilon}^1 = -\frac12 \Gamma^{0123} \epsilon, \qquad\tilde{\epsilon}^2 = -\frac12 \epsilon + \zeta\,,
\ee
the transformations can be written as
\ba\label{eq:eps}
\delta_\epsilon  \phi^I_r &=& \frac12\bar\lambda\Gamma^I\left[\mathbbm{1}+\beta\right] \epsilon +\xi^\mu_\epsilon \partial_\mu \phi^I_r\,, \cr
\delta_\epsilon  \lambda &=&- \frac1{2}\left[\mathbbm{1}  - \beta\right]\epsilon+\xi^\mu_\epsilon\partial_{\mu}\lambda\,, \cr
 \delta_\epsilon  A_\mu&=&-\frac12\bar \lambda\big(  \Gamma_\mu + \Gamma_I\partial_\mu\phi^I_r\big) \left[\mathbbm{1} + \beta\right]\epsilon +\frac12 \bar \lambda \Gamma_m\left[ \tfrac{1}{3}\mathbbm{1}  +\beta\right] \epsilon\ \bar\lambda \Gamma^m \partial_\mu \lambda+\xi^\nu_\epsilon F_{\nu\mu} \,,\qquad
\ea
and
\ba \label{eq:zeta}
\delta_{\zeta} \phi^I_r &=& -{\bar\lambda}\Gamma^I\zeta +\xi^\mu_\zeta\partial_\mu \phi^I_r\,, \cr
\delta_{\zeta} \lambda &=& \zeta +\xi^\mu_\zeta\partial_{\mu}\lambda\,, \cr
 \delta_{\zeta} A_\mu&=&\bar\lambda \big(  \Gamma_\mu + \Gamma_I\partial_\mu\phi^I_r\big)\zeta -\frac{1}{3} \bar\lambda \Gamma_m \zeta \ \bar\lambda \Gamma^m \partial_\mu \lambda +\xi^\nu_\zeta F_{\nu\mu}\,,
\ea
where
\ba\label{eq:beta}
\xi^\mu_\epsilon &=&-\frac12 \bar \lambda \Gamma^\mu [\mathbbm{1}+\beta] \epsilon\,,\qquad \xi^\mu_\zeta =\bar \lambda \Gamma^\mu  \zeta \,,\cr
\beta&=&-\rmi\cG \lp \mathbbm{1} +\frac{1}{2} \hat \Gamma^{\mu\nu} {\cal F_{\mu\nu}} +\frac18 \hat{\Gamma}^{\mu_1\nu_1\mu_2\nu_2} {\cal F}_{\mu_1\nu_1} {\cal F}_{\mu_2\nu_2} \rp \Gamma _{(0)}^{D3}\Gamma _*^{(3)}\,,\cr
\Gamma_{(0)}^{D3} &=& \frac{1}{4!\sqrt{|G|}}\varepsilon^{\mu_1\dots \mu_{4}}\hat \Gamma_{\mu_1\dots
\mu_{4}}\,,\qquad  \hat{\Gamma}_\mu = \Pi_\mu^M \Gamma_M\,,\qquad\Gamma _*^{(3)}=-\rmi\Gamma ^{0123}\,,\cr
\cG&=&\frac{\sqrt{\left|G\right|}}{\sqrt{\left| G + \cal{F}\right|}}= \left[ \det\left( \delta _\mu {}^\nu + {\cal F}_{\mu \rho }G^{\rho \nu }\right) \right] ^{-1/2}\,.
\ea

Expanding $\beta$ as a function of the fields, one finds to leading order that $\beta = 1 + \ldots$. The transformations given in eqn.\ \eqref{eq:eps} then look like linear supersymmetry transformation that are deformed by higher order contributions from the expansion of $\beta$. The transformations in eqn.\ \eqref{eq:zeta} however are non-linear and these supersymmetries are spontaneously broken since the fermion $\lambda$ transforms as $\delta_\zeta \lambda = \zeta + \ldots$.

\subsection{Identifying the non-linear transformations}

Due to the constant shift in $\delta_\zeta \lambda$ this symmetry is non-linear, however, so is any combination of the $\zeta$-transformation and the $\epsilon$-transformation. So the question is how we identify the correct non-linearly realized supersymmetries on the D3-brane? A simple way to accomplish this is to recall some basic facts about the standard O3-plane projection. This projection preserves the sixteen supercharges that are linearly realized on the D3-brane and projects out the 16 supercharges that are non-linearly realized on the D3-brane. The projection constrains the spinors to satisfy
\be
\tilde{\epsilon}^1 = \Gamma_{456789} \tilde{\epsilon}^2 = - \Gamma_{0123} \tilde{\epsilon}^2.
\ee
If we now plug in the spinor redefinitions from eqn. \eqref{eq:spinorredef} into the above equation, we find
\ba
\zeta = 0\,.
\ea
As expected the O3 orientifold projection preserves only the linearly realized SUSY transformations for a D3-brane, which are generated by $\epsilon$ (since our equation above implies that the non-linear trafos generated by $\zeta$ are absent). So the linear transformations are given by eqn. \eqref{eq:eps}.

After this trivial check we can now identify the correct non-linear supersymmetries. The sixteen supercharges that would be  projected out by an O3-orientifold projection are given by a spinor that satisfies
\be
\tilde{\epsilon}^1 = -\Gamma_{456789} \tilde{\epsilon}^2 = + \Gamma_{0123} \tilde{\epsilon}^2.
\ee
These are the 16 supersymmetries that are non-linearly realized on a D3-brane. Let us again plug in the redefinitions from equation \eqref{eq:spinorredef} and we find
\be
\epsilon =  \zeta\,.
\ee
So we find that the correct non-linear transformations are given by the combination of the $\epsilon$ and $\zeta$ transformations with $\epsilon=\zeta$.

Having identified the correct linear and non-linear supersymmetries on the D3-brane we can now spell out the explicit transformation laws. Note that these transformations are not unique, in the sense that we can change them by redefining the worldvolume fields. This usually allows one to simplify the transformations substantially and bring them into a standard form. We  show this explicitly for the non-linear transformation by expanding $\beta$ to next to next to leading order. After field redefinitions we find (to the order we are working in) the standard non-linear transformations laws one expects.\footnote{The expansion of $\beta$ in terms of the worldvolume fields leads to an infinite series, so the field redefinitions likewise do not terminate at any finite order in the fields.} 

The explicit expansion of $\beta$ as defined in \eqref{eq:beta} is given to a certain subleading order in appendix \ref{sec:beta}. Using this result, we find for the non-linear supersymmetry transformations with $\zeta=\epsilon$
\ba\label{eq:trafos}
(\delta_\epsilon +\delta_\zeta)\phi^I_r &=& \frac12 \bar \lambda \Gamma^I \lp \frac12 \Gamma^{\mu\nu} F_{\mu\nu}+\partial^\mu \phi^{J}_r\Gamma_{J\mu} \rp \epsilon+\ldots \cr
&=& \frac12 (\bar \lambda \Gamma^{\mu} \epsilon) \partial_\mu \phi^{I}_r+\frac14 (\bar \lambda  \Gamma^{I\mu\nu} \epsilon) F_{\mu\nu}+\frac12 (\bar \lambda \Gamma^{IJ\mu} \epsilon) \partial_\mu \phi_{r,J}+ \ldots\,,\cr
(\delta_\epsilon +\delta_\zeta) \lambda &=&\epsilon - \frac12 (\bar \lambda \Gamma_{M} \partial_{\mu} \lambda) \Gamma^{M\mu}\epsilon+\frac14 F_{\mu\nu}\Gamma^{\mu\nu}\epsilon +\frac12 \partial^\mu \phi^I_r \Gamma_{I\mu} \epsilon-\frac18 F^{\mu\nu}F_{\mu\nu} +\frac18 F_{\mu\nu} \tilde{F}^{\mu\nu}\Gamma_{0123} \epsilon\cr
&&-\frac 14\partial_\mu \phi^I_r \partial^\mu \phi_{r,I}  \epsilon -\frac14 \partial^\mu \phi^{I}_r  \partial^\nu \phi^{J}_r \Gamma_{IJ\mu\nu} \epsilon +\frac14 F^{\mu\nu} \partial^\sigma \phi^{I}_r\Gamma_{I\mu\nu\sigma}\epsilon+\ldots \,,\cr
(\delta_\epsilon +\delta_\zeta) A_\mu &=& -\frac12 \bar \lambda \Gamma_\mu \lp \frac12 F_{\rho\sigma}\Gamma^{\rho\sigma} + \partial^\rho \phi^I_r \Gamma_{I\rho}\rp\epsilon+\ldots\cr
 &=& \frac12 (\bar \lambda \Gamma^\nu \epsilon) F_{\nu\mu}+\frac12 (\bar \lambda \Gamma_{I}\epsilon) \partial_\mu \phi^I_r-\frac12 (\bar \lambda \Gamma_{I\rho\mu}\epsilon) \partial^\rho \phi^I_r -\frac14  (\bar \lambda \Gamma_{\mu\rho\sigma}\epsilon) F^{\rho\sigma}+\ldots\,.
\ea
In the transformations of the scalars and the vectors, we recognize the first term as the standard non-linear transformation, see \eqref{eq:nonlinear_app} and appendix \ref{app:nonlinear} for more information on non-linear realizations. For the spinor, such a term is however absent. We will massage the purely fermionic term in the transformation to obtain the standard transformation. For three 10d spinors of the same chirality we can use the Fierz identity
\be
\Gamma^M \lambda_1 (\bar \lambda_2 \Gamma_M \lambda_3)+\Gamma^M \lambda_2 (\bar \lambda_3 \Gamma_M \lambda_1)+\Gamma^M \lambda_3 (\bar \lambda_1 \Gamma_M \lambda_2)=0\,,
\ee
to rewrite the purely fermionic term in $(\delta_\epsilon +\delta_\zeta) \lambda$ as follows
\be
- \frac12 (\bar \lambda \Gamma_{M} \partial_{\mu} \lambda) \Gamma^{M\mu}\epsilon
= \frac12 (\bar \lambda \Gamma^{\mu} \epsilon )\partial_\mu \lambda - \frac14\Big{(}\Gamma^{M\mu} \epsilon\ (\bar \lambda \Gamma_{M} \partial_{\mu} \lambda) +\Gamma^{M\mu } \lambda\ (\bar \epsilon \Gamma_{M} \partial_\mu\lambda ) \Big{)} - P(\epsilon) \Gamma^M \partial_M \lambda\,,\label{eq:redefinition}
\ee
with $P(\epsilon)$ defined as
\be
P(\epsilon) = -\frac 14 \left(\lambda \otimes \bar \epsilon -\epsilon \otimes \bar \lambda  + \Gamma^M (\bar \epsilon\Gamma_M \lambda)\right)\,.
\ee
The first term in \eqref{eq:redefinition} is the non-linear transformation we are after. The term between brackets can be absorbed into a field redefinition. The last term in \eqref{eq:redefinition}  can be compensated by a transformation that leaves the action trivially invariant (this is called a `zilch symmetry' or `trivial symmetry' \cite{Bergshoeff:2013pia}). The leading order term in such a transformation is  $\delta_P\lambda = P \Gamma^M \partial_M \lambda$, with $(C P)^T = CP$ and $C$ the charge conjugation matrix. This is an invariance of the leading term in the action $\bar \lambda \Gamma^\mu \partial_\mu \lambda$.

We can now simplify the above transformations by field redefinitions. In particular, we define
\ba
\tilde \phi^I_r &=& \phi^I_r -\frac18 (\bar \lambda  \Gamma^{I\mu\nu} \lambda) F_{\mu\nu}-\frac14 (\bar \lambda \Gamma^{IJ\mu} \lambda) \partial_\mu \phi_{r,J}\,,\cr
\tilde \lambda &=& \lambda+\frac14\Gamma^{M\mu} \lambda\ (\bar \lambda \Gamma_{M} \partial_{\mu} \lambda) -\frac14 F_{\mu\nu}\Gamma^{\mu\nu}\lambda -\frac12 \partial^\mu \phi^I_r \Gamma_{I\mu} \lambda+\frac18F^{\mu\nu}F_{\mu\nu} -\frac18 F_{\mu\nu} \tilde{F}^{\mu\nu}\Gamma_{0123} \lambda\cr
&&+\frac 14 \partial_\mu \phi^I_r \partial^\mu \phi_{r,I}  \lambda +\frac14 \partial^\mu \phi^{I}_r  \partial^\nu \phi^{J}_r \Gamma_{IJ\mu\nu} \lambda-\frac14 F^{\mu\nu} \partial^\sigma \phi^{I}_r\Gamma_{I\mu\nu\sigma}\lambda\,,\cr
\tilde A_\mu &=& A_\mu +\frac14 (\bar \lambda \Gamma_{I\rho\mu}\lambda) \partial^\rho \phi^I_r +\frac18  (\bar \lambda \Gamma_{\mu\rho\sigma}\lambda) F^{\rho\sigma}\,.
\ea
To the order we are working in this leads, after an additional rescaling of the fermions $\tilde\lambda, \epsilon \rightarrow \sqrt{2}\tilde\lambda,\sqrt{2}\epsilon$, to the simplified non-linear sixteen supersymmetry transformations
\ba\label{eq:nltrafos}
\delta  \tilde \phi^I_r \equiv (\delta_\epsilon +\delta_\zeta)\tilde \phi^I_r|_{\zeta=\epsilon} &=& (\bar{\tilde{\lambda}} \Gamma^{\mu} \epsilon)\ \partial_\mu \tilde{\phi}^{I}_r\,,\cr
\delta\tilde \lambda \equiv (\delta_\epsilon +\delta_\zeta+\delta_{P(\epsilon)}) \tilde{\lambda}|_{\zeta=\epsilon} &=&\epsilon + (\bar{\tilde{\lambda}} \Gamma^\mu \epsilon)\ \partial_\mu \lambda\,,\cr
\delta \tilde A_\mu \equiv (\delta_\epsilon +\delta_\zeta)|_{\zeta=\epsilon} \tilde{A}_\mu &=&  (\bar{\tilde{\lambda}} \Gamma^\nu \epsilon)\ \tilde F_{\nu\mu}+ (\bar{\tilde{\lambda}} \Gamma_{I} \epsilon) \partial_\mu \tilde{\phi}^I_r\,.
\ea
Note that we cannot remove any further terms via field redefinitions since $\bar{\tilde \lambda} \Gamma^M \tilde \lambda =0$, $\forall M$. 

The above transformations are consistent with applying T-duality to the standard non-linear transformations for a D9-brane. In particular, the standard transformation for a 10d vector field
\be
\delta A_M = (\bar \lambda \Gamma^N \epsilon) F_{NM}\,,
\ee
gives after T-duality, $A_M \rightarrow \{A_\mu(\sigma^\mu), -\phi^I_r(\sigma^\mu)\}$, the following transformations for the scalar $\phi^I_r$ and the vector $A_\mu$ 
\ba
\delta \phi_{r,I} &=& (\bar \lambda \Gamma^\nu \epsilon) \partial_\nu \phi_{r,I}\,,\cr
\delta A_\mu &=& (\bar \lambda \Gamma^\nu \epsilon) F_{\nu\mu} + (\bar \lambda \Gamma^I \epsilon) \partial_{\mu} \phi_{r,I} \,.
\ea
Since this argument does not rely on the particular range of $\mu$ and $I$ we conclude that the non-linearly transformations that we derived in eqn. \eqref{eq:nltrafos} for a D3-brane are the same for any D$p$-brane.

We can also reduce the above transformation laws in eqn. \eqref{eq:nltrafos} to four dimensions by decomposing the 10d spinor $\tilde \lambda$ into four 4d Majorana spinors $\tilde \lambda^0$ and $\tilde \lambda^i$, $i=1,2,3$, where $\tilde \lambda^0$ is a singlet and $\tilde \lambda^i$ is a triplet under the SU(3) symmetry acting on the three complex transverse directions. We also switch to complex scalars $\tilde{\phi}^i= \tilde \phi_r^{2i-1}+\rmi \tilde \phi_r^{2i}$. We are particularly interested in the 4d $\N=1$ supersymmetry generated by $\epsilon^0$, the SU(3) singlet component of the 10d spinor $\epsilon=\zeta$. This is the non-linearly realized $\N=1$ that is preserved for example in $CY_3$ compactifications. Restricting to these four supercharges and switching to 4d $\gamma^\mu$ matrices, we find the following 4d $\N=1$ non-linear supersymmetry transformations
\ba
\delta_{\epsilon^0}\tilde \phi^i &=& (\bar{\tilde{\lambda}}^0 \gamma^\mu \epsilon^0) \partial_\mu \tilde \phi^i\,,\cr
\delta_{\epsilon^0} \tilde{\lambda}^0 &=&\epsilon^0 + (\bar{\tilde{\lambda}}^0 \gamma^\mu \epsilon^0) \partial_\mu \tilde \lambda^0 \,,\cr
\delta_{\epsilon^0} \tilde{\lambda}^i &=&(\bar{\tilde{\lambda}}^0 \gamma^\mu \epsilon^0) \partial_\mu \tilde \lambda^i \,,\cr
\delta_{\epsilon^0} \tilde{A}_\mu &=& (\bar{\tilde{\lambda}}^0 \gamma^\nu \epsilon^0) \tilde F_{\nu\mu}-\frac{1}{\sqrt{2}} \delta_{i\ib} \ls (\bar{\tilde \lambda}_+^i \epsilon_+^0) \partial_\mu \bar{\tilde \phi}^\ib + (\bar{\tilde \lambda}_-^\ib \epsilon_-) \partial_\mu \tilde \phi^i\rs\,,
\ea
where we used $\pm$ to denote 4d spinors that satisfy $\lambda_\pm = \frac12 (1 \pm \gamma_*) \lambda$, where $\gamma_* = - \rmi \gamma^{0123}$. After an additional field redefinition $\tilde A_\mu \rightarrow \tilde A_\mu +\frac{1}{\sqrt{2}} \delta_{i\ib} \ls (\bar{\tilde \lambda}_+^i \tilde \lambda_+) \partial_\mu \bar{\tilde \phi}^\ib + (\bar{\tilde \lambda}_-^\ib \tilde \lambda_-) \partial_\mu \tilde \phi^i\rs$ we find the standard non-linear 4d $\N=1$ supersymmetry transformations
\ba\label{eq:nltrafos2}
\delta_{\epsilon^0}\tilde \phi^i &=& (\bar{\tilde{\lambda}}^0 \gamma^\mu \epsilon^0) \partial_\mu \tilde \phi^i\,,\cr
\delta_{\epsilon^0} \tilde{\lambda}^0 &=&\epsilon^0 + (\bar{\tilde{\lambda}}^0 \gamma^\mu \epsilon^0) \partial_\mu \tilde\lambda^0 \,,\cr
\delta_{\epsilon^0} \tilde{\lambda}^i &=&(\bar{\tilde{\lambda}}^0 \gamma^\mu \epsilon^0) \partial_\mu\tilde \lambda^i \,,\cr
\delta_{\epsilon^0} \tilde{A}_\mu &=& (\bar{\tilde{\lambda}}^0 \gamma^\nu \epsilon^0) \tilde F_{\nu\mu}\,.
\ea
Note again that field redefinitions could not remove a term like $(\bar{\tilde{\lambda}}^0 \gamma^\mu \epsilon^0) \partial_\mu \tilde \phi^i$ since $\bar{\tilde{\lambda}}^0 \gamma^\mu \tilde \lambda^0=0$. So it is a very non-trivial check that the transformation laws take this form after appropriate field redefinitions. It provides strong support for your conjecture since fields that transform as above can be packaged into the constrained superfields in eqn. \eqref{eq:constraints}, as we review in appendix \ref{app:nonlinear}. 

\subsection{A different \texorpdfstring{$\kappa$}{}-symmetry gauge fixing}
In the subsection above we chose a particular simple $\kappa$-symmetry gauge fixing to derive the general non-linear supersymmetry transformations for the worldvolume fields on a D3-brane in flat space. The corresponding non-linear transformations for an anti-D3-brane in flat space are identical. These general transformations given in eqn. \eqref{eq:nltrafos2} are certainly consistent with the truncation discussed above in subsection \eqref{ss:only_SU(3)singlets}, where one only keeps the Goldstino $\tilde{\lambda}^0$ and the vector $\tilde{A}_\mu$. As discussed above such a truncation arises in the KKLT setup where the triplets $\tilde{\phi}^i$ and $\tilde{\lambda}^i$ get a mass. However, for the other truncations we need to place the D3-brane or anti-D3-brane on top of orientifolds to remove some of the worldvolume fields. In this case the $\kappa$-symmetry fixing has to be compatible with the orientifold projection and we can normally not use the above $\kappa$-symmetry gauge fixing. The particular case of an anti-D3-brane on top of an O3-plane, discussed in subsection \ref{ss:only_fermions}, was worked out in \cite{Kallosh:2014wsa}. So here we  work out the truncation discussed in subsection \ref{ss:only_SU(3)triplets}.

We can use $X^M = \{ X^m, \phi^I_r \}$ in the general transformation laws given in eqn. \eqref{eq:branetrafos} and rewrite them as
\ba\label{eq:trafos2}
\delta \theta &=& \epsilon + (\mathbbm{1} + \Gamma)\kappa+\xi^\mu \partial_\mu \theta\,,\cr
\delta X^m &=& -\bar{\theta} \Gamma^m \epsilon + \bar{\theta} \Gamma^m (\mathbbm{1} + \Gamma) \kappa + \xi^\mu \partial_\mu X^m\,,\cr
\delta \phi_r^I &=& -\bar{\theta} \Gamma^I \epsilon + \bar{\theta} \Gamma^I (\mathbbm{1} + \Gamma) \kappa + \xi^\mu \partial_\mu \phi_r^I\,,\cr
\delta A_\mu &=& -\bar{\theta} \Gamma_M \sigma_3 \epsilon \partial_\mu X^M + \frac16 \bar{\theta} \sigma_3 \Gamma_M \epsilon \bar{\theta} \Gamma^M \partial_\mu \theta + \frac16 \bar{\theta} \Gamma_M \epsilon \bar{\theta} \sigma_3 \Gamma^M \partial_\mu \theta + \bar{\theta} \sigma_3 \Gamma_M (\mathbbm{1}+\Gamma)\kappa\ \partial_\mu X^M \cr
&&  -\frac12 \bar{\theta} \sigma_3 \Gamma_M (\mathbbm{1}+\Gamma) \kappa \ \bar{\theta} \Gamma^M \partial_\mu \theta -\frac12 \bar{\theta} \Gamma_M (\mathbbm{1} + \Gamma)\kappa \ \bar{\theta}\sigma_3 \Gamma^M \partial_\mu \theta + \xi^\nu F_{\nu\mu} \,.
\ea

We take the anti-D3-brane to extend along the $0123$ directions and fix the diffeomorphism invariance $X^m = \delta_\mu^m \sigma^\mu$. Next we do two $O7$ orientifold projections that remove the complex scalars $\phi^2=\phi_r^3+\rmi \phi_r^4$, $\phi^3 =\phi_r^5+\rmi \phi_r^6$ and the gauge field $A_\mu$. Since the anti-D3-brane is placed on top of the O7-planes all spinor doublets $f=(f^1,f^2)$, i.e. the background supersymmetries $\epsilon$ as well as the worldvolume spinor $\theta$ and $\kappa$, have to satisfy
\be\label{eq:Oprojections}
f^1 = -\Gamma_{01234567} f^2=-\Gamma_{01234589} f^2\,.
\ee
Each of these conditions removes half the supercharges so that we are left with 8 supercharges, half of which are linearly and half of which are non-linearly realized, so that we have $\N=1+1$ in 4d.

The $\Gamma$-matrix that appears in the $\kappa$-symmetry (cf. eqn. \eqref{eq:branetrafos}) takes in this case the form 
\ba
\Gamma &=& \rmi \sigma_2 \Gamma_{(0)}^{D3}\cr
&=&\rmi \sigma_2 \Gamma^{0123}\lp 1 + (\partial^\mu \phi^L_r-\bar \theta \Gamma^L\partial^\mu \theta)\Gamma_{L\mu}-\frac12 \partial_\mu \phi^L_r \partial^\mu \phi_{r,L} -\frac12 \partial^\mu \phi^{L_1}_r \partial^\nu \phi^{L_2}_r \Gamma_{L_1L_2\mu\nu}+\ldots\rp,\qquad\quad\label{eq:Gamma}
\ea 
where $L=1,2$ and $\ldots$ denotes higher order terms. We choose the $\kappa$-gauge fixing to be 
\be\label{eq:kfix}
(\mathbbm{1}+\rmi \sigma_2 \Gamma^{0123}) \theta = 0\qquad \Rightarrow \qquad \theta_1 = \Gamma_{0123} \theta_2\,.
\ee
The above orientifold projections in eqn. \eqref{eq:Oprojections} together with this $\kappa$-symmetry gauge fixing removes all but four components of the worldvolume spinors. The remaining four components correspond to the SU(3) singlet spinor $\lambda^0$. 

We now work out the explicit transformations for the fields to the relevant order. The orientifold truncation preserves for each 16 component 10d MW doublet two 4d spinors that satisfy 
\be
(1 \pm\rmi \sigma_2 \Gamma^{0123})f_\pm=0\,,\qquad f_\pm = \frac 12 (1 \mp\rmi \sigma_2 \Gamma^{0123})f\,.
\ee
Here $f_+$ corresponds to the SU(3) singlet and $f_-$ to the spinor with the 1 index (the same index as $\phi^1$). We have gauged away the worldvolume field $\theta_-$ corresponding to $\lambda^1$ and kept $\theta_+$ corresponding to $\lambda^0$. $\theta_+$ transforms non-linearly under $\epsilon_+$ and linearly under $\epsilon_-$. We are interested in the non-linear symmetries generated by $\epsilon_+$, which in particular means we can set $\epsilon_-=0$. 

In order to preserve our $\kappa$-symmetry gauge fixing in eqn. \eqref{eq:kfix} we have to demand that (cf.\ eqn. \eqref{eq:trafos2})
\be\label{eq:thetam}
0=\delta \theta_- = [(\mathbbm{1} + \Gamma)\kappa]_-\,,
\ee
where we used $\epsilon_-=0$ and $\bar \theta \Gamma^L\partial^\mu \theta= \bar \theta_+ \Gamma^L\partial^\mu \theta_-+\bar \theta_- \Gamma^L\partial^\mu \theta_+=0$ for $\theta_-=0$. To leading order we find, using \eqref{eq:Gamma}
\be
\kappa_- = -\frac12 \partial^\mu \phi^L_r\Gamma_{L\mu} \kappa_+ + \ldots\,.
\ee
This relation dramatically simplifies the leading order result for the other chirality of $(\mathbbm{1} + \Gamma)\kappa$
\be
[(\mathbbm{1} + \Gamma)\kappa]_+ = 0\,.
\ee
This then leads to the expected fermion transformation
\ba
\delta \theta_+ =\epsilon_+ +\xi^\mu \partial_\mu \theta_++\ldots\,.
\ea
The parameter $\xi^\mu$ can be obtained by demanding the preservation of the diffeomorphism gauge fixing $\delta X^m = 0$
\be
\xi^\mu = \bar \theta_+ \Gamma^\mu \epsilon_+ +\ldots\,,
\ee
where we used that $\bar \theta (\mathbbm{1}+\Gamma)\kappa = \bar \theta_+[ (\mathbbm{1}+\Gamma)\kappa]_+$.

Now let us look at the transformation of the $\phi^I_r$
\be
\delta \phi_r^I =\bar{\theta}_+ \Gamma^I [(\mathbbm{1}+\Gamma)\kappa]_- + \xi^\mu \partial_\mu \phi^I_r=  \xi^\mu \partial_\mu \phi^I_r\,,
\ee 
where we used eqn. \eqref{eq:thetam}. The above transformation is the expected one for $\phi^1 =\phi_r^1 + \rmi \phi_r^2$ and is consistent with the vanishing of $\phi^3_r, \phi^4_r, \phi^5_r$ and $\phi^6_r$. 

Lastly we can check $\delta A_\mu$. We find by explicit calculation that $\delta A_\mu = \xi^\nu F_{\mu\nu}$, which is consistent with $A_\mu=0$.

To summarize, after reducing the spinors to four dimensions and dropping the in this case irrelevant sub- and superscripts, we get the following transformation rules for the single 4d fermion $\lambda$ and the one complex scalar $\phi$
\ba
\delta \lambda &=& \epsilon +( \bar \lambda \gamma^\mu \epsilon) \partial_\mu \lambda\,,\cr
\delta \phi &=&( \bar \lambda \gamma^\mu \epsilon) \partial_\mu \phi\,.
\ea

So again we find the expected non-linear transformations that match the transformations of the constrained chiral multiplets $S$ and $H$ after the field redefinitions discussed in appendix \ref{app:nonlinear}.

\section{Discussion}\label{sec:conclusion}

D-branes play an important role in string phenomenology. They have been used to construct (semi-) realistic models of particle physic and are also often used in string models of inflation and the construction of dS vacua. Any realistic string model of our universe requires the breaking of supersymmetry. If supersymmetry is broken by (anti-) D-branes, then this breaking is spontaneous and we should be able to write down a 4d supersymmetric action.\footnote{The constraints on the interactions of the Goldstino with other fields, that arise from the non-linearly realized supersymmetry, are discussed for intersecting D-brane models in \cite{Antoniadis:2004uk}.}  Here we have taken an important step in this direction by working out the explicit non-linear supersymmetry transformations for the worldvolume fields of an (anti-) D3-brane. We also provided evidence that the worldvolume fields, the vector $A_\mu$, the three complex scalars $\phi^i$ and the four spinors $\lambda^0$, $\lambda^i$, can be packaged into the 4d $\N=1$ superfields $W_\alpha$, $H^i$, $S$ and $Y^i$ which satisfy the constraints $S^2=S W_\alpha = S \bar{D}_{\dot \alpha} \bar H^\ib=S Y^i=0$. We hope this allows for detailed future studies of the phenomenological aspects of string vacua with spontaneously broken supersymmetry of the type performed in \cite{Antoniadis:2004uk, Choi:2005ge} .

For cosmology we have to be able to promote the global supersymmetry of the constrained multiplets from the D3-brane to a local supersymmetry, since we have to solve Einstein's equations in order to get the cosmological evolution. Fortunately, this problem was solved recently for rather general constraints on multiplets in \cite{Ferrara:2016een,Dall'Agata:2016yof}, in some details for de Sitter supergravity in \cite{Bergshoeff:2015tra} and for the  Born-Infeld multiplet in \cite{Aoki:2016cnw}.

We have shown here that all for cosmology interesting constrained multiplets can arise in simple string theory constructions, including the real orthogonal multiplet $\Phi$ that satisfies the constraint $S(\Phi-\bar \Phi)=0$. Therefore we believe that  the importance of such cosmological models is significant, beyond the successful  phenomenological applications to cosmology,  which they represent.

\section*{Acknowledgments}

We are grateful to I.\ Antoniadis, E.\ Dudas, I.\ Garc\'ia-Etxebarria, S.\ Kachru, A.\ Linde, L.\ McAllister, H. P.\ Nilles, J.\ Polchinski, S.\ Parameswaran, E.\ Silverstein, F.\ Quevedo, A.\ Uranga, A.\ Van Proeyen and I.\ Zavala for enlightening discussions and to our collaborators on related projects, E.\ Bergshoeff, J.\ J.\ Carrasco, K.\ Dasgupta,  S.\ Ferrara, F.\ Quevedo and J.\ Thaler which made the current project possible to accomplish.   The work of RK is supported by the SITP, and by the NSF Grant PHY-1316699.   The work of TW is supported by COST MP1210. BV acknowledges support from a Starting Grant  of the European Research Council (ERC  STG  Grant  279617). We thank the organizers of STRING PHENO 2016 for hospitality during the final stage of this project.

\appendix

\section{Expanding \texorpdfstring{$\beta$}{}}\label{sec:beta}
In this appendix we expand $\beta$ as defined in eqn. \eqref{eq:beta} up to terms of dimension four. We restrict to the $\kappa$ and diffeomorphism fixing described in subsection \ref{sec:fixsimple} (see also appendix A.2 in \cite{Bergshoeff:2013pia}).

For the determinants we use the expansion valid for $(4\times 4)$-matrices $M$
\be
\det (\mathbbm{1}+M) = 1 + {\rm Tr}\,  M + \frac 12( ({\rm Tr}\,M)^2- {\rm Tr} ( M^2)) + \ldots\,.
\ee
For the metric we have 
\be
-\text{det}(G) = 1 +\partial_\mu \phi_r^I \partial^\mu \phi_{r,I}-2\bar \lambda \Gamma^\mu \partial_\mu \lambda+\ldots\,.
\ee
This then gives
\be
\frac{1}{\sqrt{-\text{det}(G)}} = 1 - \frac12 \partial_\mu \phi_r^I \partial^\mu \phi_{r,I}+ \bar \lambda \Gamma^\mu \partial_\mu \lambda+\ldots \,.
\ee
Next we look at ${\cal G}$
\ba
{\cal G} &=& \frac{\sqrt{-\text{det}(G)}}{\sqrt{-\text{det}(G+{\cal F})}} = \ls \text{det} \lp \delta_\mu^\nu + {\cal F}_{\mu\rho} G^{\rho \nu} \rp \rs^{-\frac12} = \ls 1 + {\cal F}_{\mu\rho} G^{\rho \mu} +\frac12  \mathcal{F}_{\mu\nu} \mathcal{F}^{\mu\nu}+ \ldots \rs^{-\frac12}\cr
&=&  \ls 1  +\frac12 F^{\mu\nu}F_{\mu\nu}+ \ldots \rs^{-\frac12} = 1-\frac14 F^{\mu\nu}F_{\mu\nu}+ \ldots\,.
\ea
From these two expansions above, it is clear that the expansion of $\beta$ will not terminate but rather lead to ever higher powers of subsonic fields that will appear in the transformation laws. We expect that these can be removed via field redefinitions as is the case for the lower dimensional terms.

We also need
\ba
\frac 12 \hat{\Gamma}^{\mu\nu} {\cal F}_{\mu\nu}
&=&\frac 12  \ls \Gamma^{\mu\nu}+2\partial^{[\mu} \phi_{r}^I \Gamma_I{}^{\nu]}+\ldots \rs \lp F_{\mu\nu} +2 \bar \lambda \Gamma_{[\nu} \partial_{\mu]} \lambda +\ldots\rp\cr
&=& \frac 12  \Gamma^{\mu\nu}F_{\mu\nu}  +F^{\mu\nu} \partial_\mu \phi_r^I \Gamma_{I\nu} - \Gamma^{\mu\nu} \bar \lambda \Gamma_{\mu} \partial_{\nu} \lambda+\ldots
\ea
and
\ba
\frac 18 \hat{\Gamma}^{\mu_1\nu_1\mu_2\nu_2} {\cal F}_{\mu_1\nu_1} {\cal F}_{\mu_2\nu_2} =\frac 1 8 \Gamma^{\mu_1\nu_1\mu_2\nu_2} F_{\mu_1\nu_1} F_{\mu_2\nu_2} + \ldots =\frac 14 \Gamma_{0123} F_{\mu\nu} \tilde{F}^{\mu\nu} + \ldots
\ea
where $\tilde{F}^{\mu\nu} = \tfrac12\epsilon^{\mu\nu\rho\sigma} F_{\rho \sigma}$. 

Lastly we need
\ba
\Gamma_{(0)}^{D3} &=&\frac 1 {4!\sqrt{-\det G}} \varepsilon^{\mu_1 \ldots \mu_4}\hat \Gamma_{\mu_1 \ldots \mu_4}\cr
&=&\frac 1 {4!}  \varepsilon^{\mu_1 \ldots \mu_4}\frac 1 {\sqrt{-\det G}}
\Pi_{\mu_1}^{m_1}\Pi_{\mu_2}^{m_2}\Pi_{\mu_3}^{m_3}\Pi_{\mu_4}^{m_4}  \Gamma_{m_1\ldots m_4}
\cr
&=&\frac 1 {4!}\varepsilon^{\mu_1 \mu_2 \mu_3 \mu_4}( 1 - \frac 12 \partial_\mu \phi^I \partial^\mu \phi_I + \bar \lambda \Gamma^{\mu}  \partial_{\mu} \lambda+\ldots)\Big{(}\Gamma_{\mu_1 \mu_2 \mu_3 \mu_4} +4(\partial_{\mu_!} \phi^I-\bar \lambda \Gamma^{I}\partial_{\mu_1} \lambda )  \Gamma_{I\mu_2\mu_3\mu_4}\cr
&&-4(\bar \lambda \Gamma^{\nu}\partial_{\mu_1} \lambda )  \Gamma_{\nu\mu_2\mu_3\mu_4}+6\partial_{\mu_1} \phi^I \partial_{\mu_2} \phi^J\Gamma_{IJ\mu_3\mu_4}\Big{)}\cr
&=& -\Gamma_{0123} \left(1   +( \partial^{\mu} \phi_r^I   - \bar \lambda \Gamma^I \partial^\mu \lambda)\Gamma_{I\mu}- \frac 12 \partial_\mu \phi^I_r \partial^\mu \phi_{r,I}-\frac 12  \partial^{\mu} \phi_r^I \partial^{\nu} \phi_r^J\Gamma_{IJ\mu \nu}+\ldots\right)
\ea
where we used $\epsilon^{0123}=-1$ and the identities
\be
\Gamma_{\mu\nu} \Gamma_{0123} =\frac 12 \epsilon_{\mu\nu\rho\sigma} \Gamma^{\rho\sigma}\,,\qquad
\Gamma_{\mu} \Gamma_{0123} = -\frac 1{3!} \epsilon_{\mu\nu\rho\sigma} \Gamma^{\nu\rho\sigma}\,.
\ee

Now we put everything together to get from eqn. \eqref{eq:beta} the following expansion
\ba
\beta &=& {\cal G} \lp \mathbbm{1} +\frac{1}{2} \hat \Gamma^{\mu\nu} {\cal F_{\mu\nu}} +\frac18 \hat{\Gamma}^{\mu_1\nu_1\mu_2\nu_2} {\cal F}_{\mu_1\nu_1} {\cal F}_{\mu_2\nu_2} \rp \Gamma^{D3}_{(0)} \Gamma_{0123} \cr
&=& 1 -  (\bar \lambda \Gamma_{M} \partial_{\mu} \lambda) \Gamma^{M\mu}+\frac12 \Gamma^{\mu\nu} F_{\mu\nu}+\partial^\mu \phi_r^{I}\Gamma_{I\mu}-\frac12 \partial_\mu \phi_r^I \partial^\mu \phi_{r,I} -\frac14 F^{\mu\nu}F_{\mu\nu}  +\frac14 F_{\mu\nu} \tilde{F}^{\mu\nu}\Gamma_{0123}\cr
&& -\frac12 \partial^\mu \phi_r^{I}  \partial^\nu \phi_r^{J} \Gamma_{IJ\mu\nu}+\frac12 F^{\mu\nu} \partial^\sigma \phi_r^{I}\Gamma_{I\mu\nu\sigma}+ \ldots\,.
\ea

\section{Non-linear realizations and constrained multiplets}\label{app:nonlinear}

We give a brief review of constrained superfields and non-linear realizations of supersymmetry. In short, the constraints act as effective ways to reduce the components of multiplets. The remaining components can have a non-linearly realized supersymmetry.

The constraints we advocate in eqn.\ \eqref{eq:constraints}, offer a convenient way of putting every worldvolume field in a different constrained multiplet. To map the field content on the brane (with transformations \eqref{eq:nltrafos2}) to the fields in the constrained multiplets, one needs to perform a field redefinition that we spell out below.

\subsection{Non-linear realization of supersymmetry}
It is important to note how one arranges the components of superfields to transform nonlinearly under broken supersymmetries, even before using constraints. The VA realization on the Goldstino of broken supersymmetry is
\be
\delta \lambda = \epsilon + (\bar \lambda \Gamma^\mu \epsilon) \partial_\mu \lambda\,,\label{eq:goldstino_app}
\ee
and matter fields (scalars, vectors, fermions \ldots) transform as
\be
\delta_\epsilon \phi =  (\bar \lambda \Gamma^\mu \epsilon) \partial_\mu \phi\,.\label{eq:nonlinear_app}
\ee

This follows from the discussion of Coleman, Wess and Zumino on non-linear realizations of a broken bosonic symmetry group \cite{Coleman:1969sm} and the extension to broken supersymmetry \cite{rocek}. The results can be paraphrased as \emph{linear representations (multiplets) of a group can be decomposed into direct sums of non-linearly transforming fields, by using a group transformation with the Goldstone field acting as the group parameter.} This means that for any superfield $\Phi$, one can define a non-linearly transforming version $\hat \Phi$ as:
\be
\hat \Phi(x,\theta) \equiv \Phi (x',\theta') \equiv \exp(\bar \lambda Q) \cdot \Phi(x,\theta)\,,\label{eq:nonlinear_superfield}
\ee
with $Q$ the generator of broken supersymmetry and $\lambda$ the Goldstino field.
As discussed in the second reference of \cite{rocek}, this transforms as
\be
\delta_\epsilon \hat \Phi =  (\bar \lambda \Gamma^\mu \epsilon) \partial_\mu \hat \Phi\,.
\ee
In particular, all the components of the superfield individually transform  as \eqref{eq:nonlinear_app}.

To describe effective Lagrangians for low mass modes, one can eliminate components of a superfield that describe very massive fields by constraints. Those constraints become exact in the limit where these massive fields become infinitely massive \cite{Komargodski:2009rz,Kallosh:2015pho}. One has to take care when identifying the correct non-linearly transforming low mass fields after the constraints have been incorporated. We illustrate this with some examples.

\subsection{Superfield constraints to remove components}

Constrained superfields are useful tools to organize the field content when supersymmetry is broken. However, one still needs to perform a field redefinition to find components that transform as in \eqref{eq:nonlinear_app}. We give these explicit field redefinitions here. We only discuss the constraints \eqref{eq:constraints} in rigid supersymmetry. A more general approach to properties of constrained superfields coupled to gravity was developed recently in \cite{Ferrara:2016een}. A simple classification of constrained multiplets was proposed in  \cite{Dall'Agata:2016yof}, where the authors suggest that to get rid of super-partners one can take any multiplet $Q$ and impose the constraint $S\bar S Q=0$. The constraints we discuss in this paper follow from this more general discussion. The universal formula is presented in eq.\ (6) of  \cite{Dall'Agata:2016yof}. 

 In the rest of this section we work in four-dimensions and we use the conventions of \cite{Freedman:2012zz} (see their appendix 14A for superfields). In particular we denote
\be
\gamma_* = i \gamma_{0123}\,,\qquad P_L = \frac 12 (1+ \gamma_*)\,,\qquad P_R = \frac 12 (1-\gamma_*)\,.
\ee

\subsubsection{Goldstino}

When supersymmetry is broken, we can describe the Goldstino with a nilpotent chiral superfield obeying
\be
S^2 =0\,, 
\qquad S = s + \frac 1{\sqrt 2} \bar \theta P_L G  + \frac 14 \bar \theta P_L \theta F\,.  
\ee
This constraint eliminates the boson from the spectrum, as it puts $s = G^\alpha G_\alpha/2F$. The Goldstino transforming as \eqref{eq:goldstino_app} is related to the fermion $G$ of the nilpotent field as
\be
\lambda^0(x)  \equiv \frac{G(x_-)}{\sqrt2 F(x_-)}\,, \qquad x_- = x - \frac 12 \bar \lambda^0(x) \gamma^\mu\gamma_*\lambda^0(x)\,.
\ee
For more information on this field redefinition, see for instance \cite{Samuel:1982uh} or the appendix of \cite{Vercnocke:2016fbt}.

\subsubsection{Fermions}

In addition to the nilpotent Goldstino, one can take other constrained superfields. Take for instance ``orthogonal'' chiral superfields, obeying
\begin{equation}
 S Y^i = 0\,, \qquad Y^i = y^i +\frac 1{ \sqrt 2}\bar \theta P_L \psi^i_y + \frac 14 \bar \theta P_L \theta F^i_y\,.
\end{equation}
As explained in \cite{Komargodski:2009rz}, this makes the scalars dependent fields
\begin{equation}
 y^i = \frac{\bar G P_L \psi^i_y}{F} - \frac{\bar G P_L G}{2 F^2} F_y^i\,,
\end{equation}
and hence each orthogonal multiplet describes only one fermion $\psi^i_y$.

However, the spinors $\psi^i_y$ do not by themselves transform in the standard way \eqref{eq:nonlinear_app}.
One can work out  the components of the non-linearly transforming superfields $\lambda^i$, defined through \eqref{eq:nonlinear_superfield}. 
The non-linearly transforming scalar components is zero, while the non-linearly transforming fermion $\lambda^i \equiv\hat \psi^i_y$ is implicitly defined by
\ba
\lambda^i &=&
[1+ Q(\lambda^0(x))] [\psi^i_y(x_-) - 2 F^i (x_-) \lambda^0(x)]\,,\quad Q(\lambda^0) =  2 P_L \gamma^\mu \lambda^ \otimes \partial_\mu \bar \lambda^0 P_L - 2 P_R \gamma^\mu \lambda^0 \otimes \partial_\mu \bar \lambda^0 P_R\,,\nonumber\\
x_- &=&  x - \frac 12 \bar \lambda^0(x) \gamma^\mu\gamma_*\lambda^0(x)\,.
\ea
%
For more details and notation in other superspace conventions we again refer to the appendix of \cite{Vercnocke:2016fbt}.

\subsubsection{Scalars}

To describe complex scalars, one can take chiral superfields $H^i=  h^i + \frac 1 {\sqrt2} \bar \theta P_L  \psi^i +\frac 14  \bar \theta P_L \theta F^i$ obeying the constraint
\be
S \bar D_{\dot \alpha} \bar H = 0\,.
\ee
This constraint makes the fermion a function of the (derivative of) the scalar component, see for instance \cite{Komargodski:2009rz}.

Note that the complex scalar $h$ does not transform in the standard way. The scalar that transforms according to \eqref{eq:nonlinear_app} is rather
\be
\hat h(x) \equiv h (x_+)\,,\qquad x_+ = 1 + \frac 12 \bar \lambda^0(x) \gamma^\mu\gamma_*\lambda^0(x)\,.
\ee

\subsubsection{Vector}

It might seem troublesome that the vector on the brane does not have a standard transformation of the form \eqref{eq:nonlinear_app}, but rather
\be
\delta_\mu A_\mu = \bar \lambda^0 \gamma^\nu \epsilon F_{\nu\mu}\,.
\ee
However, this can be compensated by a gauge transformation and a contraction with the VA vielbein $dx^\mu + \bar \lambda^0 \gamma^\mu d \lambda^0$. See \cite{Klein:2002vu} for a discussion on this point.

One can eliminate the gaugino from a field-strength superfield $W_\alpha$, by the constraint $S W_\alpha =0$.
The field redefinition that brings the gauge field in $W_\alpha$ to a non-linearly transforming vector $A_\mu$ can be found in \cite{Klein:2002vu} .

\end{document}